\documentclass{emulateapj}
\usepackage{natbib}
\usepackage{amssymb,amsmath}
\usepackage{hyperref}

\newcommand{\bs}{\boldsymbol}
\newcommand{\argmin}{\operatornamewithlimits{argmin}}

\begin{document}
\submitted{Accepted to ApJ Letters}
\title{Spitzer, Gaia, and the Potential of the Milky Way}

\author{Adrian~M.~Price-Whelan\altaffilmark{1,2} Kathryn V. Johnston\altaffilmark{1}}

\altaffiltext{1}{Department of Astronomy, Columbia University, 550 W 120th St., New York, NY 027, USA}
\altaffiltext{2}{\email{adrn@astro.columbia.edu}}

\begin{abstract}

Near-future data from ESA's {\it Gaia} mission will provide precise, full phase-space information for hundreds of millions of stars out to heliocentric distances of $\sim$10 kpc. This ``horizon'' for full phase-space measurements is imposed by the {\it Gaia} parallax errors degrading to worse than 10\%, and could be significantly extended by an accurate distance indicator. Recent work has demonstrated how {\it Spitzer} observations of RR Lyrae stars can be used to make distance estimates accurate to 2\%, effectively extending the {\it Gaia}, precise-data horizon by a factor of ten in distance and a factor of 1000 in volume. This \emph{Letter} presents one approach to exploit data of such accuracy to measure the Galactic potential using small samples of stars associated with debris from satellite destruction. The method is tested with synthetic observations of 100 stars from the end point of a simulation of satellite destruction: the shape, orientation, and depth of the potential used in the simulation are recovered to within a few percent. The success of this simple test with such a small sample in a single debris stream suggests that constraints from multiple streams could be combined to examine the Galaxy's  dark matter halo in even more  detail --- a truly unique opportunity that is enabled by the combination of {\it Spitzer} and {\it Gaia} with our intimate perspective on our own Galaxy.

\end{abstract}


\section{Introduction}
\label{intro.sec}
 The existence of vast halos of unseen {\it dark} matter surrounding each galaxy has long been proposed to explain the surprisingly large
motions of the {\it baryonic} matter that we can see \citep[e.g.,][]{rubin70}.
Dark-matter-only simulations of structure formation lead us to expect that these dark matter halos should have density distributions that are described by a universal radial profile \citep{navarro96} with a variety of triaxial shapes \citep{jing02}.
The inclusion of baryons in the simulations tends to soften the triaxiality of the dark matter in the inner regions of the halo \citep[e.g., as the disk forms,][]{bailin05} and
can alter the radial profile through a combination of adiabatic contraction and energetic feedback \citep[e.g.][]{pontzen12}.
Hence, measurements of the shape, orientation, radial profile, and extent of dark matter halos provides information about the formation of these vast structures, as well as the messy baryonic processes that continue to shape them.

The Milky Way is the best candidate for such a detailed study of a dark matter halo
since we can resolve large samples of stellar tracers.
Thousands of blue horizontal branch stars selected from the Sloan Digital Sky Survey (SDSS) have been used to probe the Milky Way mass out to tens of
kpc \citep[SDSS, see][]{deason12a,kafle12}, and estimates with combined tracers extend to 150kpc \citep{deason12b}.

This approach assumes that the tracers represent a random sampling of phase-mixed orbits drawn from a smooth distribution function, however large area surveys have revealed the existence of large-scale spatial inhomogeneities in the form of giant stellar streams \citep{newberg02,majewski03,belokurov06}, demonstrating that a significant fraction of the stellar halo is neither randomly sampled nor is fully phase-mixed.

A complimentary approach to measuring the mass distribution is to instead take advantage of the {\it non}-random nature of the Galaxy's stellar distribution and utilize the knowledge that stars in streams were once all part of the same object.
Such approaches can require orders of magnitude fewer tracers than a randomly sampled population to achieve comparable accuracy.
One method is to simply fit orbits to observations of streams 
\citep[e.g.,][]{koposov10}.
However, the assumption that debris traces a single orbit is actually incorrect \citep[see][]{johnston98,helmi99} and
changes in orbital properties along debris streams can lead to systematic biases in measurements of the Galactic potential \citep{eyre09a,varghese11}.
\citet{sanders13a} recently demonstrated that this bias is equally problematic for the very thinnest, coldest streams, whose observed properties may be indistinguishable from those of the parent orbit \citep[e.g. such as the globular cluster, GD1 --- see][]{koposov10}, as for the much more extended and hotter streams \citep[e.g. such as debris from the Sagittarius dwarf galaxy --- see][]{majewski03} where offsets from a single orbit are clearly apparent.

One way to address these biases is to run self-consistent N-body simulations of satellite destruction in a variety of potentials with the aim of simultaneously constraining both the properties of the satellite and the Milky Way.
Many studies of the Sagittarius debris system (hereafter Sgr) have adopted this approach,
with the most recent work attempting to place constraints on the triaxiality and orientation of the dark matter halo
\citep{law10}.

The promise of near-future data sets including full phase-space information has also inspired other approaches. \citet{binney08} and \citet{penarrubia12} demonstrate that the distribution of energy and entropy in debris, respectively, will be minimized only for a correct
assumption of the form of the Galactic potential.
\citet{sanders13b} examine the distribution of debris in action-angle co-ordinates and show that stars stripped from the same disrupted object must lie along a single line in angle-frequency space, providing a constraint that can be used as a potential measure.

In this \emph{Letter} we re-examine and update a complimentary approach to using tidal debris as a potential measure \citep[originally proposed by][]{johnston99a}  in the context of current and near-future observational capabilities, and apply it to a simulation of the Sgr debris system.
In Section \ref{sec:context} we outline the observational prospects and Sgr properties that motivated this re-examination.
In Section \ref{sec:method} we present the updated potential measure and test it with synthetic observations of simulated Sgr debris.
In Section \ref{sec:discussion} we highlight the advantages and shortcomings of this method.
We conclude in Section \ref{sec:conclusion}.

\section{Context and motivation} \label{sec:context}
The method presented in Section \ref{sec:method} takes advantage of
three distinct developments: (i)
the demonstration of a technique for deriving distances to individual
RR Lyrae stars with 2\% accuracies (Section \ref{sec:spitzer}); (ii)
the prospect of proper motion measurements of the same stars with
$\sim$10~$\mu$as/yr precision (Section \ref{sec:gaia}); and (iii) the
tracing of debris associated with Sgr around the entire Galaxy
(Section \ref{sec:sgr})

\subsection{{\it Spitzer} and 2\% distance errors to RR Lyrae in the halo}
\label{sec:spitzer}

There is a long tradition for using RR Lyrae stars in the Galaxy to
study structure 
\citep[e.g.][]{shapley18}, substructure
\citep[e.g.][]{sesar10}, and distances to satellite galaxies
\citep[e.g.][]{clementini03}.  However, studies of RR Lyrae at optical
wavelengths are limited by both metallicity effects on the intrinsic
brightness of these stars and variable extinction along the line of
sight.  Moreover, systematic differences between instruments make it
difficult to tie observations across the sky to a common scale. 

At longer wavelengths, RR Lyrae promise tighter constraints on
distances.  
\citet{madore12} have recently shown, 
using five stars with
trigonometric parallaxes measured by Hubble \citep{benedict11},
that the dispersion in the mid-IR Period-Luminosity (PL) relation 
\citep[first mapped by][]{longmore86}
at
wavelengths measurable by NASA's {\it Spitzer} mission is $\sim$0.03~mag.
This implies that it is
possible to use {\it Spitzer} to determine distances that are good to $2\%$ for 
individual RR Lyrae stars out to $\sim$60~kpc ({\it Spitzer}'s limit for detecting and measuring RR Lyrae).
For comparison, distance measurements of Blue Horizontal Branch
stars typically
achieve $\sim$10-15\% uncertainties  \citep[if appropriate color measurements are available, e.g.,][]{deason12b}.

\begin{figure}[h!]
\begin{center}
\includegraphics[width=0.45\textwidth]{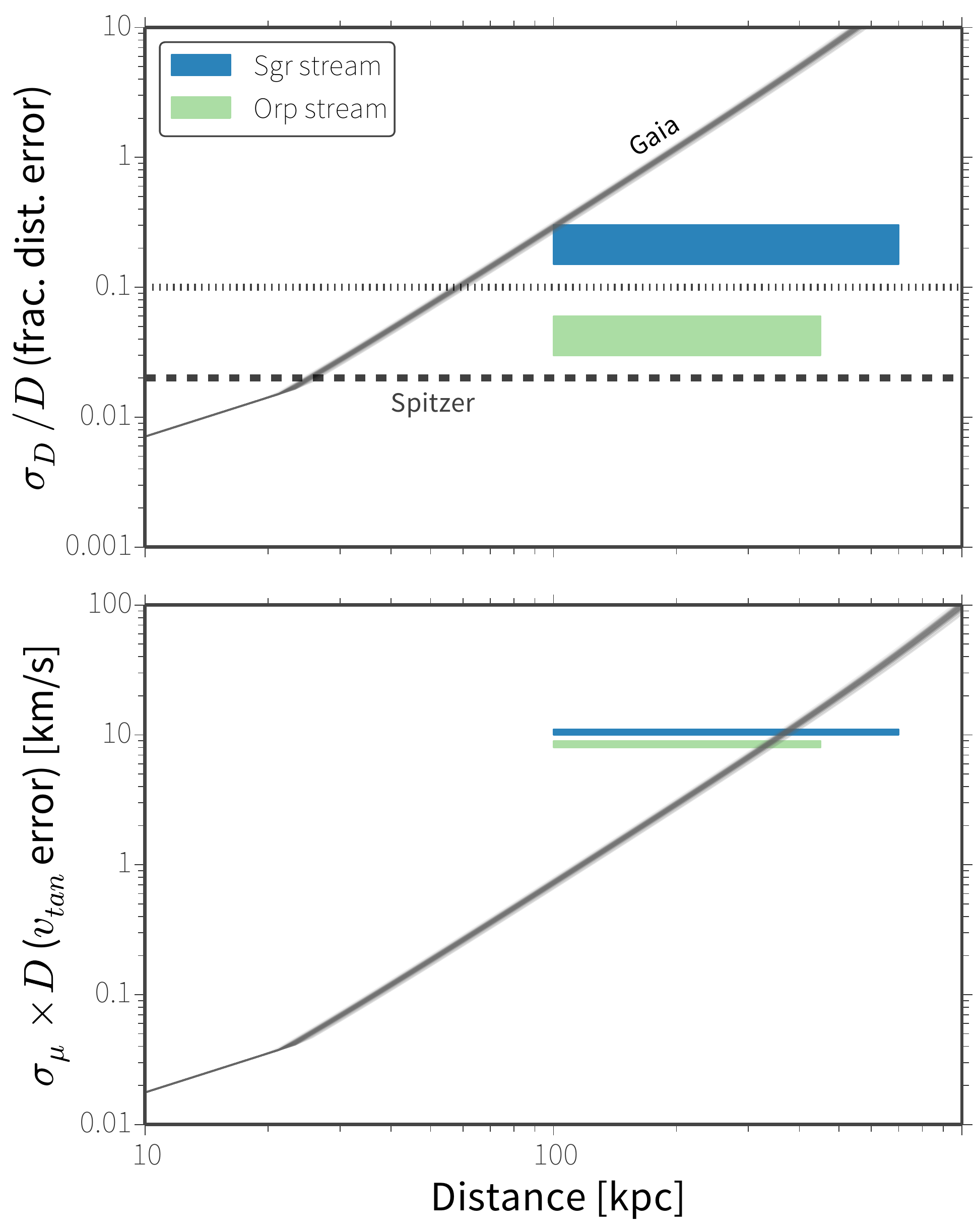}

\caption{Expected {\it Gaia} distance and tangential velocity errors as a function of heliocentric distance for RR Lyrae stars. Errors are a function of color and magnitude of the source, and hence the metallicity: each line is computed
by Monte Carlo sampling from the empirical metallicity distribution of
the Galactic halo from \cite{ivezic08}. Parallax distance errors from {\it Gaia} are larger than the line-of-sight size of both Sgr and Orphan (Orp), but photometric distance errors are comparable to the the Sgr scale (assuming 10\% errors, dotted line). Bottom panel shows that the {\it Gaia} tangential velocity errors are smaller than the internal velocity dispersion of nearer regions of both Sgr and Orp. }\label{fig:gaia_errors}
\end{center}
\end{figure}

\subsection{Gaia and the age of astrometry}
\label{sec:gaia}
The {\it Gaia} satellite \citep{gaia01} is
an astrometric mission which aims to measure the positions of billions
of stars with 10-100~$\mu$as accuracies. Combined with expected 
proper motion accuracies, this will enable full six-dimensional phase-space 
maps of the Galaxy with $<$10\% distance errors for heliocentric distances of
up to $\sim$6~kpc for RR Lyrae stars.

Figure~\ref{fig:gaia_errors} shows the {\it Gaia} end-of-mission distance
and tangential velocity error estimates for RR Lyrae. Within 2~kpc, {\it Gaia} will measure distances to these stars with
better than 2\% accuracy --- RR Lyrae in this volume can be used to
test and calibrate the {\it Spitzer} PL relation described above. Beyond the
2~kpc threshold, the mid-IR PL relation for RR Lyrae will provide better 
distance measurements. 

The combination of {\it Spitzer} and {\it Gaia} data will 
extend the ``horizon'' of where precise, six-dimensional phase-space 
maps of the Galaxy are possible from $<$10~kpc to 60~kpc. This enormous 
increase in volume will greatly refine data on debris systems in the halo.

\subsection{The Sagittarius debris system}
\label{sec:sgr}
Sgr was discovered serendipitously during a radial velocity 
survey of the Galactic bulge \citep{ibata94}. 
Signatures of extensive stellar
streams associated with Sgr have since  been
mapped across the sky in carbon stars \citep{totten98}, M giants
selected from 2MASS \citep{majewski03}, main
sequence turnoff stars from SDSS
\citep{belokurov06}, and RR Lyrae in the Catalina Sky Survey
\citep{drake13}. 

\begin{figure}[h]
\begin{center}
\includegraphics[width=0.45\textwidth]{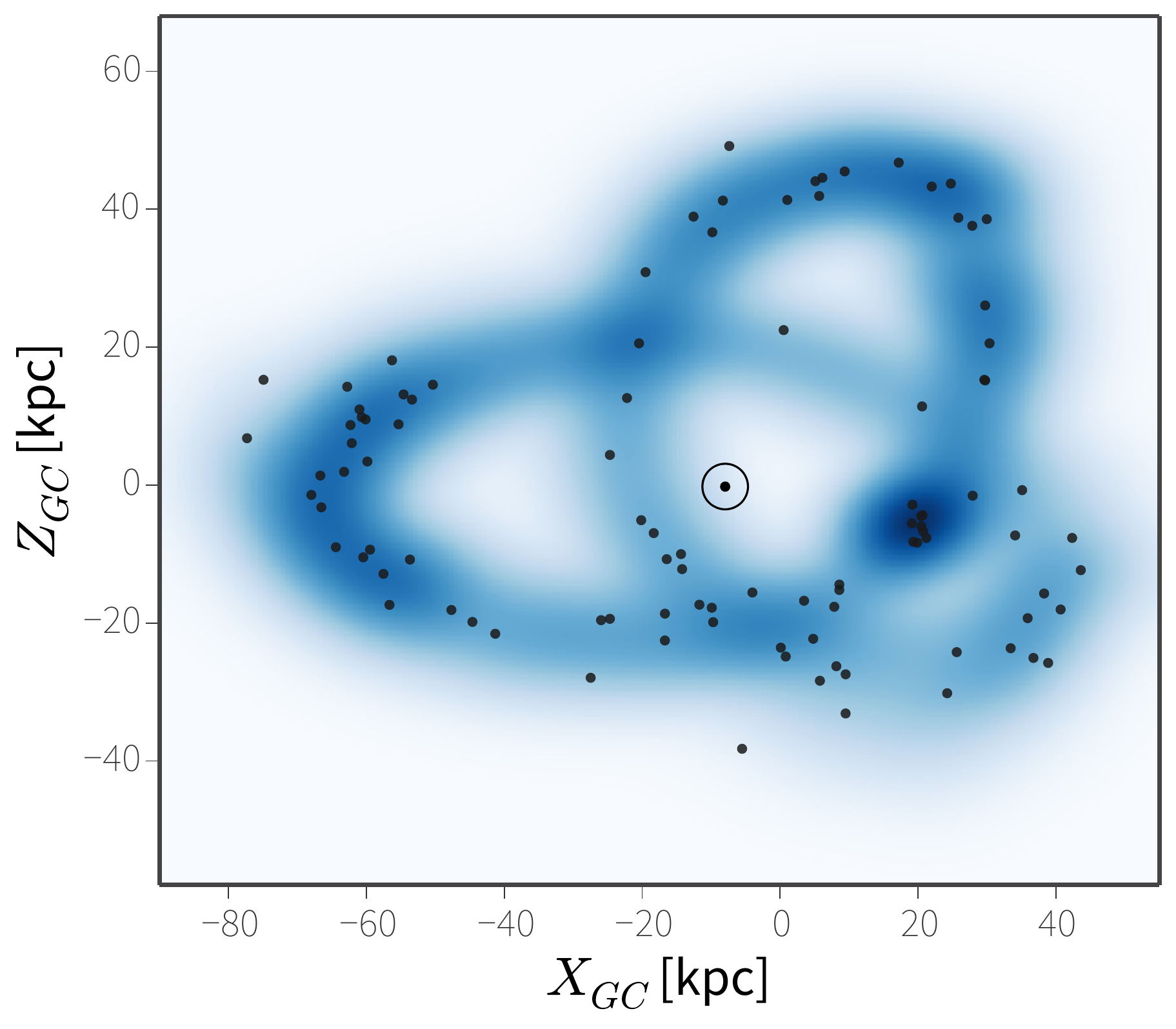}
\caption{ Particle density (blue) of the first leading and trailing wraps from the final time-step of the \citet{law10} simulation of the Sgr stream. Point markers (black) show positions of a random sample of 100 stars drawn from this density distribution. The position of the Sun is shown with the solar symbol. }\label{fig:lm10}
\end{center}
\end{figure}

Sgr stream data has inspired a rich set of  models 
\citep[e.g.,][]{johnston99b, fellhauer06}.
Most recently, \citet[][hereafter LM10]{law10} combined all
the (then) current data on the Sgr debris to constrain both a model of its evolution
and the potential in which it orbits. (Note that new observational work by 
\citet{belokurov13} suggest that the trailing tail of Sgr debris does not
match the LM10 model.)
Figure \ref{fig:lm10} shows particle positions
from the final time-step of the LM10 N-body
simulation of dwarf satellite disruption along the expected Sgr orbit
in the best-fitting Milky Way halo model. The simulation was
run in a three-component potential, with a triaxial, logarithmic halo
model of the form
\begin{equation}
  \Phi_{halo} = v_{halo}^2 \ln(C_1 x^2 + C_2 y^2 + C_3 xy + (z/q_z)^2 + R_c^2)
\end{equation}
where $C_1$, $C_2$, and $C_3$ are combinations of the $x$ and $y$ axis
ratios ($q_1$, $q_2$) and orientation of the halo with respect to the
baryonic disk ($\phi$):
\begin{align}
  C_1 &= \frac{\cos^2\phi}{q_1^2} + \frac{\sin^2\phi}{q_2^2}\\
  C_2 &= \frac{\sin^2\phi}{q_1^2} + \frac{\cos^2\phi}{q_2^2}\\
  C_3 &= 2\sin\phi\cos\phi \left(q_1^{-2} - q_2^{-2}\right).
\end{align}

A comparison of simulations and data enabled LM10 to make an assessment of the three-dimensional 
mass distribution of the Milky Way's dark matter
halo through constraints on the potential parameters $v_{\rm halo}$, $q_1$, $q_z$, and $\phi$. Combined {\it Spitzer} and {\it Gaia} measurements of distances and proper motions
of RR Lyrae in the Sgr debris will open up new avenues for potential constraints. Figure~\ref{fig:gaia_errors} shows that a 2\% distance error is smaller than the 
distance range in the stream (top panel). Similarly, {\it Gaia} proper motion
error estimates correspond to tangential velocity errors less than the velocity
dispersion for much of the stream (bottom panel). The next section outlines a 
new method to take advantage of this information. 


\section{Description and test of our algorithm}
\label{sec:method}
\begin{figure}[h]
\begin{center}
\includegraphics[width=0.47\textwidth,trim=5 0 0 0, clip]{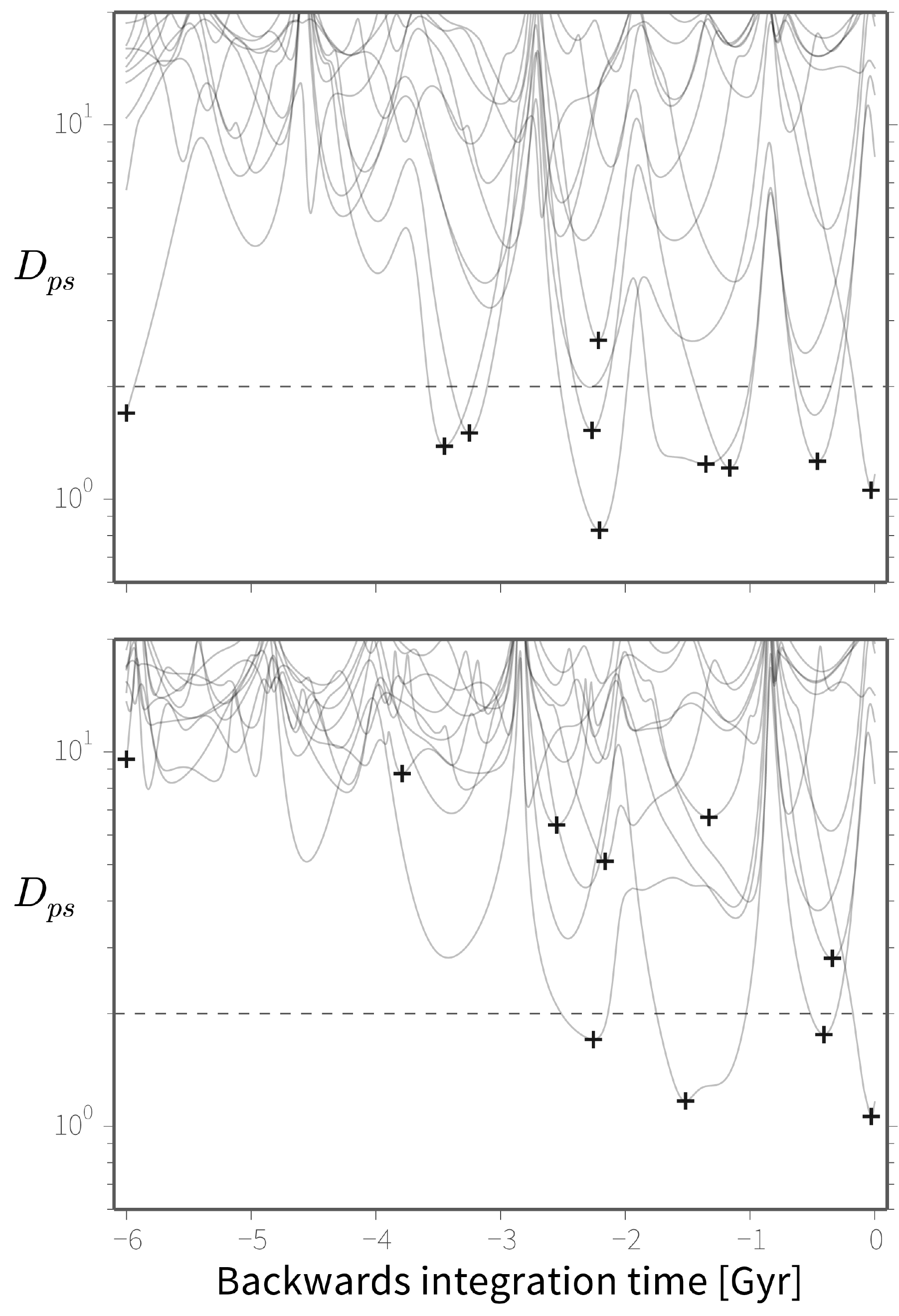}
\caption{Phase space distance ($D_{\rm ps}$) for 10 randomly selected stars integrated backwards in the correct potential (top) and a potential where $q_z$ is 25\% larger (bottom). The same 10 particles are used in both figures, so the initial conditions are identical. Horizontal (dashed) line shows $D_{\rm ps}=2$, for reference.}\label{fig:ps_distance}
\end{center}
\end{figure}

With access to 6D information for stars in a tidal
stream, each star becomes a powerful potential
measure by exploiting the fact that the stars must have come from the
same progenitor: if the orbits of the stars and progenitor are integrated 
\emph{backwards} in a a potential that accurately models the Milky Way, the stars
should recombine with the progenitor (imagine watching satellite destruction in ``rewind"). If the potential is incorrect,
the orbits of the stars will diverge from that of the progenitor and
thus will not be recaptured by the satellite system (Figure~\ref{fig:ps_distance}).

This approach was originally proposed by \citet{johnston99a} and was tested on the proposed characteristics of the Space
Interferometry Mission \citep{unwin08}. Below we present an updated version of the algorithm:
the promise of 2\% distances to RR Lyrae stars (see Section
\ref{sec:spitzer}) enables a direct measurement (rather than
approximate estimate, as previously assumed) of the position of a star within its debris
structure. The test statistic that quantifies how well stars recombine
with the satellite has also been rigorously redefined.

\subsection{The algorithm: Rewinder}
Quantifying this method requires a sample of stars with known full
space kinematics $(\bs{x}_{i}, \bs{v}_{i})|_{t=0}$ (e.g., measurements
of all position and velocity components for these stars \emph{today}
at $t=0$), the orbital parameters for the progenitor system
$(\bs{x}_p, \bs{v}_p)|_{t=0}$, and a functional form for the
potential, $\Phi({\boldsymbol\theta})$. For a given set of potential
parameters, $\boldsymbol\theta$, the orbits of the stars and
progenitor are integrated backwards for several Gigayears. At
each timestep $t_j$, for each particle $i$, a set of normalized,
relative phase-space coordinates are computed
\begin{equation}
  \bs{q}_{i} = \frac{\bs{x}_{i} -
    \bs{x}_{p}}{R_{\rm tide}}\,\,\,\,,\,\,\,\,\bs{p}_{i} = \frac{\bs{v}_{i} -
    \bs{v}_{p}}{v_{\rm esc}}
\end{equation}
where $(\bs{x},\bs{v})_{i}$ and $(\bs{x},\bs{v})_{p}$ are the
phase-space coordinates for the particles and progenitor,
respectively. These definitions require an estimate of the mass of the
satellite, $m_{sat}$, which, combined with the orbital radius of the
satellite, $R$, and the computed enclosed mass of the potential within
$R$, $M_{enc}$, sets the instantaneous tidal radius and escape
velocity,
\begin{equation}
  R_{\rm tide}=R\Big(\frac{m_{\rm sat}}{3M_{\rm enc}}\Big)^{1/3}\,\,\,\,,\,\,\,\,
  v_{\rm esc}=\sqrt{\frac{2Gm_{\rm sat}}{R_{\rm tide}}}.
\end{equation}
These quantities are computed at each time step to take the time 
dependence into account, neglecting mass-loss from the satellite.
Qualitatively, when the distance in this normalized six-dimensional
space, $D_{{\rm ps},i}=\sqrt{|\bs{q}_{i}|^2+|\bs{p}_{i}|^2} \lesssim 2$, 
the star is likely recaptured by the satellite (in the absence of errors,
 we find that $\sim$90\% of the initially bound particles come within
 this limit when integrating all orbits backwards). \citet{johnston99a} 
 imposed a similar condition as a hard boundary and maximized the 
number of recaptured particles in a given backwards-integration.
What follows is a description of an updated procedure with a 
statistically-motivated choice for an objective function.

For each star, $i$,
the phase-space distance, $D_{\rm ps}$, is computed at each timestep
$t_{j}$, and the vector with the minimum phase-space distance is stored
\begin{align}
  t^*_{i} &= \argmin_{t} D_{{\rm ps},i}\\
  \bs{A}_{i} &= (\bs{q}_{i}(t^*_{i}),\bs{p}_{i}(t^*_{i})).
\end{align}
Thus, the matrix $\bs{A}_{ik}$ contains these minimum phase-space
distance vectors for each star, where $k\in[1,6]$. Intuitively, the
variance of the distribution of minimum phase-space vectors will be
larger for orbits integrated in an incorrect potential relative to the
distribution computed from the `true' orbital history of the stars: in
an incorrect potential, the orbits of the stars relative to the orbit of 
the progenitor spread out in phase space. Thus, the \emph{generalized
 variance} of the distribution --- computed for a given set of
potential parameters, $\bs{\theta}$ --- is a natural choice for the
scalar objective function, $f(\bs{\theta})$, used in constraining the potential of the
Milky Way
\begin{align}
  \Sigma_n &= \mathrm{Cov}( \bs{A}_{ik}) \\
  f(\bs{\theta}) &= \ln \det \Sigma_n.
\end{align}

\subsection{Application to Simulated Data} \label{sec:results}
The LM10 simulation data (see Section \ref{sec:sgr}) is a perfect
test-bed for evaluating the effectiveness of this method. We start by
extracting both particle data and the satellite orbital parameters
from the present-day snapshot of the simulation
data.\footnote{\url{www.astro.virginia.edu/~srm4n/Sgr/data.html}} We
then ``observe'' a sample of 100 stars from the first leading and
trailing wraps of the stream. The radial velocity and distance errors are drawn 
from Gaussians ($\varepsilon_{\rm RV} \sim \mathcal{N}(\mu=0,\sigma=10~{\rm
  km/s})$ and $\varepsilon_{\rm D} \sim \mathcal{N}(0,0.02\times
D)$) and the proper motion errors are computed from the expected {\it Gaia} error curve.\footnote{\url{http://www.rssd.esa.int/index.php?page=Science_Performance&project=GAIA}} 

\begin{figure}[h]
\begin{center}
\includegraphics[width=0.45\textwidth]{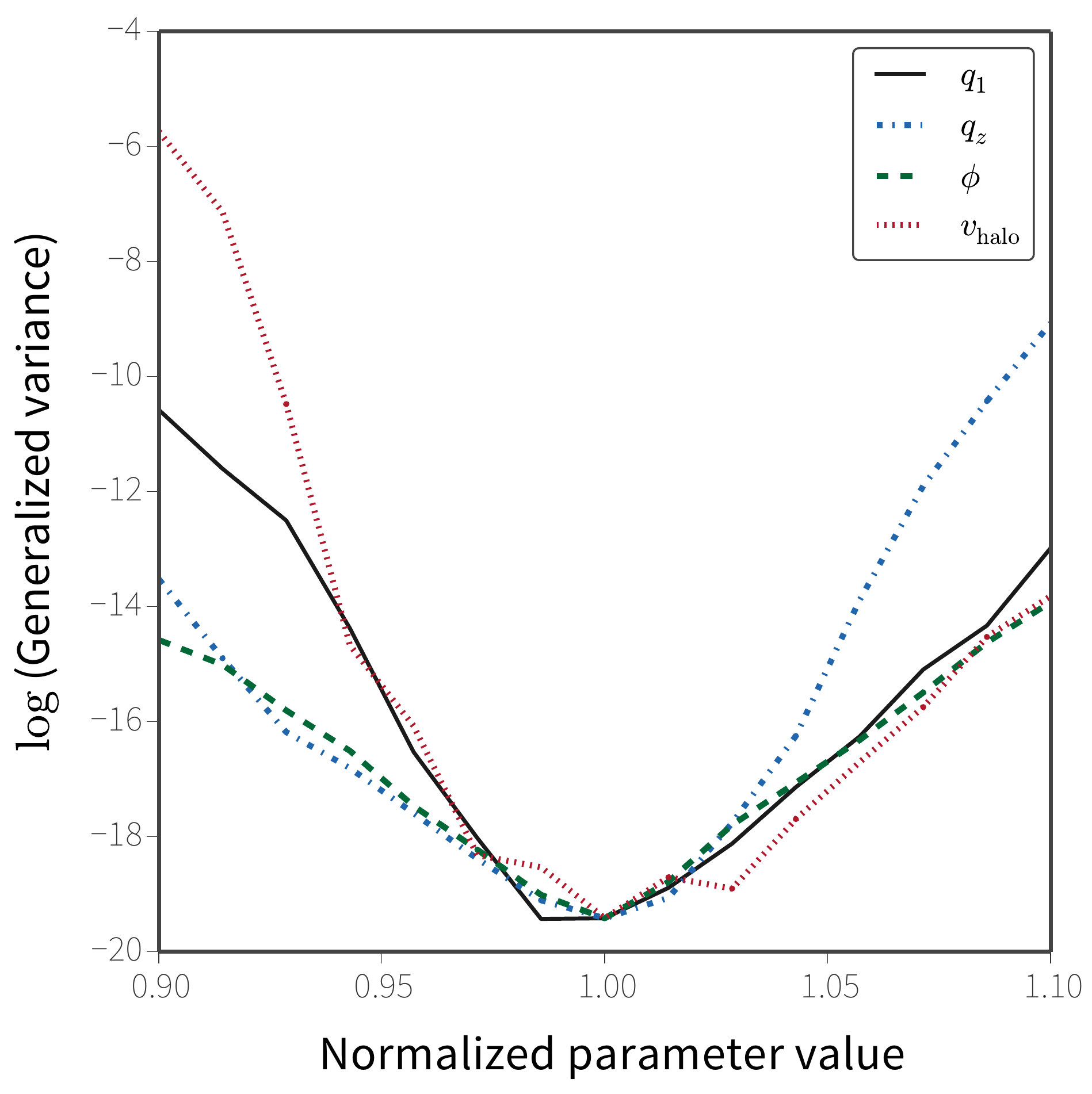}
\caption{ 1D slices of the objective function (generalized variance) for each halo potential parameter. The parameter values are normalized by the true values show the effect of varying each parameter by $\pm10$\%. The values of the objective function (vertical axis) are not interesting but note the minima around the truth (1.0).}\label{fig:objective}
\end{center}
\end{figure}

The generalized variance defines a convex function over
which we optimize four of the six logarithmic potential parameters:
$v_{circ}$, $\phi$, $q_1$, and $q_z$ ($q_2$ and $R_c$
are degenerate with combinations of the other parameters). 
Figure~\ref{fig:objective} shows one-dimensional slices of
the objective function produced by varying each of the potential
parameters by $\pm10\%$ around the true values and holding all others
fixed.

\begin{figure*}[ht!]
\centering\includegraphics[width=0.9\textwidth,trim=0 0 0 0, clip]{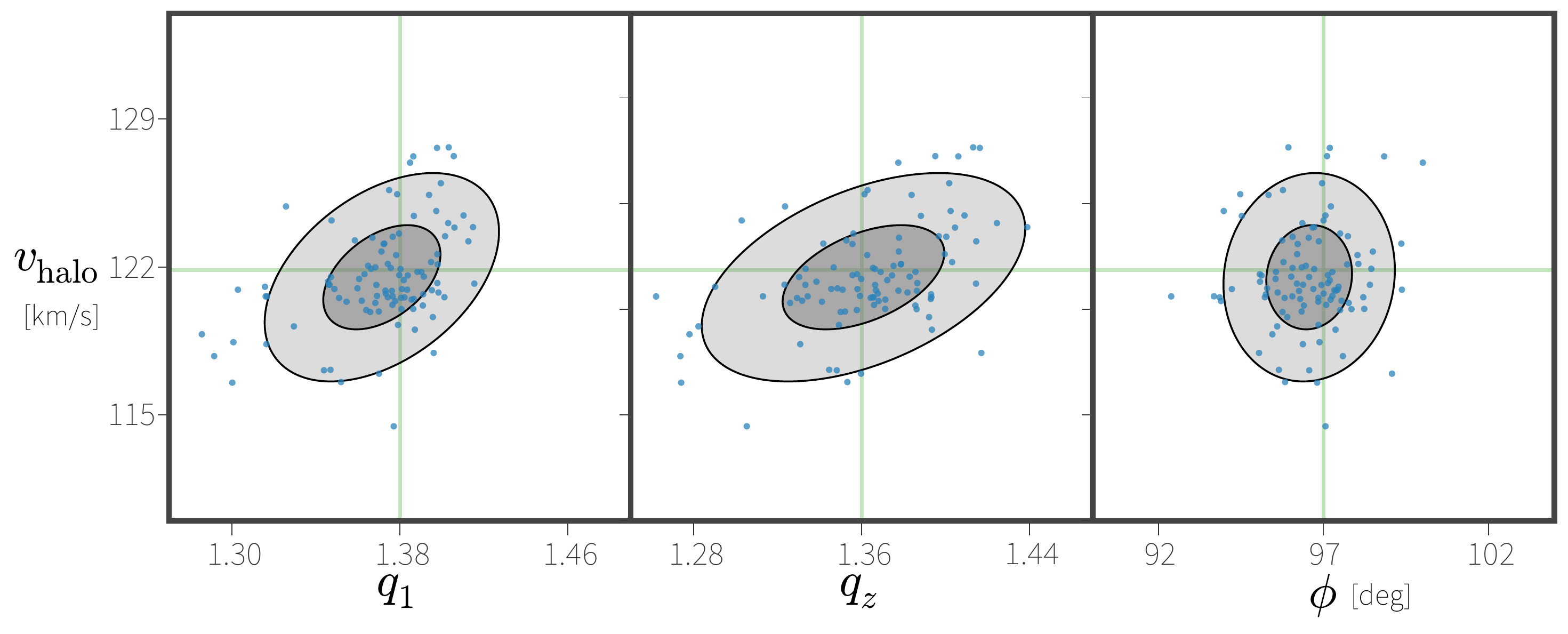}
\caption{ Blue points show the ``best-fit'' parameters resulting from each resample of 100 stars from the Sgr stream particle density shown in Figure~\ref{fig:lm10}. Green (vertical and horizontal) lines show the true values of the parameters. Grey ellipses show one- and two-sigma margins, assuming the points are normally distributed. }\label{fig:bootstrap}
\end{figure*}	

In anticipation of extending the above method to include a true
likelihood function, we use a parallelized Markov Chain Monte Carlo
(MCMC) algorithm \citep{foremanmackey2013} to sample from our
objective function.\footnote{Though MCMC is typically not an efficient
  optimization tool, in this case objective function is both noisy and
  expensive to compute. The stochasticity and easy parallelization of
  the algorithm outperforms other optimizers on this problem.}
We use the median value of the converged sample distribution as a
point estimate for the potential parameters. To assess the uncertainty
in the derived halo parameters, we sample 100 stars
100 times and estimate the potential parameters with each
resampling. Figure~\ref{fig:bootstrap} shows the recovered parameters
for each sample and demonstrates the power of this method:
a moderately sized sample of RR Lyrae
alone places strong constraints on the shape and mass of the
Galaxy's dark matter halo. From the covariance matrix derived from the 
distribution of points in Figure~\ref{fig:bootstrap}, we find the mean 
recovered parameters and one-sigma deviations to be 
$q_1 = 1.36 \pm 0.02$, $q_z = 1.36 \pm 0.03$, $\phi = 96.0 \pm 1.5$ degrees,
and $v_{\rm halo} = 123.2 \pm 1.6$ km/s.

\section{Discussion, strengths, and limitations}
\label{sec:discussion}

The strengths of this method stem from its simplicity: it requires
only a rough estimate of the satellite mass $m_{\rm sat}$ combined
with backwards integration of orbits. 
\emph{Rewinder} does not assume that stream stars follow
a single orbit and instead \emph{relies}
on the fact that each star is on a different orbit. There are also no assumptions made
about of the internal distribution of satellite
stars. Thus, \emph{Rewinder} is applicable to any debris that is
known to come from a single object and not restricted to the coldest
tidal streams. In principle, it could also be applied to the
vast stellar debris {\it clouds} that have been discovered 
\citep[e.g., the Triangulum-Andromeda and Hercules-Aquila
  clouds;][]{rochapinto04,belokurov06}, or even stars that have only
associations in orbital properties and do not form a coherent spatial
structure \citep[e.g.][]{helmi99}. The method trivially extends to
combining constraints from multiple debris systems at once by simply
integrating all debris from several satellites simultaneously, with
$D_{\rm ps}$ defined appropriately for each star.

It is also important to characterize the the limitations of this
method. Firstly, the measurement errors for RR Lyraes associated with
the very coldest streams \citep[e.g., the globular clusters Pal5 and
  GD1;][]{odenkirchen02,koposov10} will likely be too large to resolve
the minute differences in orbital properties between the debris and
satellite. Second, the present prescription neglects orbital evolution (e.g.,
dynamical friction) and scattering of stream stars due to the potential
of the satellite. Preliminary simulations (to be fully explored in 
forthcoming work) suggest that these two points can be neglected 
for satellite masses between $\sim$$10^7$ and $\sim$$10^9~\mathrm{M}_{\odot}$. Lastly, the 
current version of the algorithm relies on knowledge of the current 
position and velocity of the parent satellite, which may not be available 
\citep[e.g., the Orphan Stream;][]{belokurov07}. 

\section{Conclusions and motivation for future work}
\label{sec:conclusion}

This paper presents an algorithm for measuring the Galactic potential
that anticipates combined data from the {\it Spitzer} and {\it Gaia} satellite
missions which promise precise, full phase-space measurements of RR
Lyrae stars in the halo of our Galaxy. When applied to a sample of 100
stars (with realistic observational errors) drawn from the
\cite{law10} N-body simulation of the destruction of the Sgr dwarf
satellite, \emph{Rewinder} recovers the depth, shape, and orientation of the dark
matter potential to within a few percent.

While the tests presented in this paper are very simple, the accuracy of potential recovery promised by such a small sample of stars 
provides strong motivation for further theoretical work to: 1) develop a robust generative model that utilizes the concepts demonstrated by \emph{Rewinder}; 2) investigate the power of using multiple debris structures; and 3) examine how \emph{Rewinder} might work with less accurate measurements or missing dimensions. 

Our results also motivate an observational campaign with {\it Spitzer} to survey RR Lyrae stars in debris structures around the Milky Way to get precise distances to combine with near-future {\it Gaia} velocity data. 
If just 100 stars in a single stellar stream allow us to study the depth, shape, and orientation of the Milky Way potential, larger samples in multiple structures \citep[e.g., the Orphan Stream;][]{sesar13} offer the prospect of assessing these quantities as a function of Galactocentric radius. Tracing the mass in a dark matter halo with this level of detail is impossible for any other galaxy in the Universe.

\acknowledgments
We thank Barry Madore for providing the inspiration for this work. Thanks to 
David Hogg for statistical advice. Thanks also to Steve Majewski, David 
Law, and David Nidever. We also thank the anonymous referee for useful 
suggestions. 

APW is supported by a National Science Foundation Graduate Research
Fellowship under Grant No.\ 11-44155. This work was supported in part by 
the National Science Foundation under Grant No. PHYS-1066293 and the 
hospitality of the Aspen Center for Physics.

\bibliographystyle{apj}

\end{document}